\documentclass[10pt,twocolumn,showpacs,amsmath,amssymb,aps,prb]{revtex4-1}

\usepackage{graphicx}
\usepackage{dcolumn}
\usepackage{bm}
\usepackage[cp1250]{inputenc}
\usepackage[usenames,dvipsnames]{color}
\usepackage{tabularx}

 \definecolor{darkgreen}{rgb}{0,0.5,0} %
 \usepackage{color}
 \usepackage[normalem]{ulem}
 \usepackage{cancel}
             
\begin{document}

\title{H\"uckel--Hubbard-Ohno modeling of $\boldsymbol{\pi}$-bonds 
in ethene and ethyne\\
with application to trans-polyacetylene}

\author{M\'ate Tim\'ar$^1$}
\author{Gergely Barcza$^1$}
\author{Florian Gebhard$^2$}
\email{florian.gebhard@physik.uni-marburg.de}
\author{Libor Veis$^3$}
\email{libor.veis@jh-inst.cas.cz}
\author{\"Ors Legeza$^1$}
\email{legeza.ors@wigner.mta.hu}
\affiliation{$^1$Strongly Correlated Systems Lend\"ulet Research Group, 
Institute for Solid State Physics and Optics, MTA Wigner Research Centre for
Physics, P.O.\  Box 49, H-1525 Budapest, Hungary}
\affiliation{$^2$Fachbereich Physik, Philipps-Universit\"at Marburg,
D-35032 Marburg, Germany}
\affiliation{$^3$Dept.\ of Theoretical Chemistry, 
J.\ Heyrovsk\'{y} Institute of 
Physical Chemistry, Academy of Sciences of the Czech Republic, v.v.i, 
Dolej\v{s}kova 3, 18223 Prague 8, Czech Republic}
\date{\today}

\begin{abstract}%
Quantum chemistry calculations provide the potential energy 
between two carbon atoms in ethane (H$_3$C$-$CH$_3$), 
ethene (H$_2$C$=$CH$_2$), and ethyne (HC$\equiv$CH)
as a function of the atomic distance.
Based on the energy function for the $\sigma$-bond in ethane, $V_{\sigma}(r)$,
we use the H\"uckel model with Hubbard--Ohno interaction for the $\pi$~electrons
to describe the energies $V_{\sigma\pi}(r)$ and $V_{\sigma\pi\pi}(r)$
for the $\sigma\pi$ double bond in ethene and 
the $\sigma\pi\pi$ triple bond in ethyne, respectively.
The fit of the force functions shows that
the Peierls coupling can be estimated with some precision whereas the 
Hubbard-Ohno parameters are insignificant at the distances under
consideration. We apply the H\"uckel-Hubbard-Ohno model
to describe the bond lengths and the energies of elementary electronic excitations
of trans-polyacetylene, (CH)$_n$, and adjust the $\sigma$-bond 
potential for conjugated polymers.
\end{abstract}



\maketitle

\section{Introduction}

Electronic structure calculations for small molecules
can be performed with very good accuracy using quantum chemistry 
methods.~\cite{Jensen,Frenking} 
However, the computational cost rises strongly with the
number of valence electrons, and macromolecules with hundreds of 
delocalized valence electrons cannot be treated using accurate ab-initio 
quantum chemistry methods.
In such a situation, models with adjustable parameters are analyzed
to describe the ground state and excited states approximately.~\cite{Baeriswyl,Barford}

An example are $\pi$-conjugated polymers such as polyacetylene (PA)
or polydiacetylene (PDA). Typically, the backbone made of $\sigma$-bonds
is treated in the adiabatic approximation, and only the $\pi$-electrons
are included explicitly in the many-particle model.~\cite{Baeriswyl,Barford}
The electrons' itineracy is described by the H\"uckel
(tight-binding) Hamiltonian, and their Coulomb repulsion
is approximated by the Pariser-Parr-Pople (PPP) interaction.
Recently, we reproduced the optical properties of PDA using
a tight-binding model with a Hubbard-Ohno parameterization of the
PPP interaction between the $\pi$~electrons.~\cite{BarfordPDA}
However, the applicability of H\"uckel-PPP models
for $\pi$-conjugated systems has been put into question~\cite{SSHdebate}
because the PPP model only includes density-type interactions
and ignores bond-charge repulsion terms that could be important
for polymers; for a thorough discussion, see 
Refs.~[\onlinecite{Baeriswyl},\onlinecite{Gammel}].

Another point at issue in conjugated polymers
is the size of the `spring constant'~$K_{\sigma}^{\rm PA}$
that parameterizes the strength of the $\sigma$-bond.
In Ref.~[\onlinecite{smallKinPA}],
the value $K_{\sigma}^{{\rm PA},1}=31\, {\rm eV}/\hbox{\AA}{}^2$
was proposed for PA, close to the value for the carbon-carbon bond in ethane, 
$D_1^{\sigma,{\rm exp}}=27.2\, {\rm eV}/\hbox{\AA}{}^2$
at bond length $r_{\sigma}^{\rm exp}=1.536\, \hbox{\AA}$.~\cite{Herzberg,Helvoort}
In contrast, in Ref.~[\onlinecite{largeKinPA}] a substantially larger value,
$K_{\sigma}^{{\rm PA},2}=46\, {\rm eV}/\hbox{\AA}{}^2$,
was put forward, closer to empirical values for benzene,
$K_{\sigma}^{\rm ben}=41.3\, {\rm eV}/\hbox{\AA}{}^2$
with bond length $r_0=1.4\, \hbox{\AA}$.~\cite{benzene}  
The larger value is related
to the tendency of bonds to become stiffer at smaller atomic distances,~\cite{Mele}
and the average bond length in polyacetylene 
is the same as that in benzene.~\cite{NMR}

\begin{figure}[b]
\begin{center}
\includegraphics[width=\columnwidth]{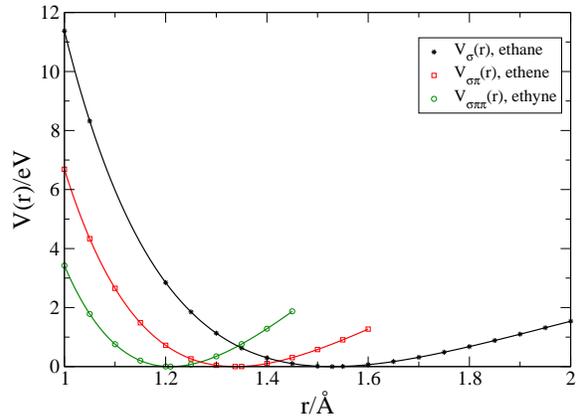}
\end{center}
\caption{(Color online) 
Ground-state energy in electron volts of 
ethane (H$_3$C$-$CH$_3$), $V_{\sigma}(r)$,
ethene (H$_2$C$=$CH$_2$, ethylene), $V_{\sigma\pi}(r)$,
and ethyne (HC$\equiv$CH, acetylene), $V_{\sigma\pi\pi}(r)$,
as a function of the carbon-carbon distance~$r$ in units of~\AA.
The energies are shifted by their values at the energy minimum.
The carbon-hydrogen bonds are fixed at their positions 
for the equilibrium structure.\label{fig:potentials}}
\end{figure}

\begin{table*}[t]
\begin{tabular}{@{}cc@{}}
(a) & (b) \\[3pt]
\begin{tabular}{|c|c|c|}
\hline
\hspace{4pt}\begin{tabular}[t]{@{\extracolsep{\fill}}ll@{}}
\multicolumn{2}{c}{H$_3$C$-$CH$_3$\vphantom{\LARGE{A}}}\\[3pt]
$r$/\AA & $V_{\sigma}(r)/{\rm eV}$\\
\hline\\[-9pt]
1.0000 &       10.8084\\
1.0500 &       7.89061\\
\mbox{}& \\
\mbox{}& \\
1.2000 &       2.68283\\
\mbox{}& \\
1.2500 &       1.74053\\
1.3000 &       1.06070\\
\mbox{}& \\
1.3500 &       0.58770\\
1.4000 &       0.27742\\
1.4500 &       0.09475\\
1.5000 &       0.01163\\
1.5290 & 0\\
1.5500 &       0.00563\\
1.6000 &       0.05877\\
1.6500 &       0.15668\\
1.7000 &       0.28788\\
1.7500 &       0.44322\\
1.8000 &       0.61542\\
1.8500 &       0.79872\\
1.9000 &       0.98856\\
1.9500 &       1.18131\\
2.0000 &       1.37414
\end{tabular}
\hspace{2pt}&\hspace{2pt}
\begin{tabular}[t]{@{}ll@{}}
\multicolumn{2}{c}{H$_2$C$=$CH$_2$}\\[3pt]
$r$/\AA & $V_{\sigma\pi}(r)/{\rm eV}$\\
\hline\\[-9pt]
1.0000 &       6.56351\\
1.0500 &       4.25500\\
1.1000 &       2.60229\\
1.1500 &       1.45774\\
1.2000 &       0.70626\\
\mbox{}& \\
1.2500 &       0.25796\\
1.3000 &       0.04249\\
1.3372 & 0\\
1.3500 &       0.00464\\
1.4000 &       0.10095\\
1.4500 &       0.29714\\
1.5000 &       0.56612\\
\mbox{}& \\
1.5500 &       0.88651\\
1.6000 &       1.24145
\end{tabular}
\hspace{2pt}&\hspace{2pt}
\begin{tabular}[t]{@{}ll@{}}
\multicolumn{2}{c}{HC$\equiv$CH}\\[3pt]
$r$/\AA & $V_{\sigma\pi\pi}(r)/{\rm eV}$\\
\hline\\[-9pt]
1.0000 &         3.42482\\
1.0500 &         1.78346\\
1.1000 &         0.75783\\
1.1500 &         0.20269\\
1.2000 &         0.00484\\
1.2097 &  0\\
1.2500 &         0.07592\\
1.3000 &         0.34680\\
\mbox{}& \\
1.3500 &         0.76329\\
1.4000 &         1.28284\\
1.4500 &         1.87198
\end{tabular}\\
\hline
\end{tabular}&
\begin{tabular}{|c|c|c|}
\hline
\hspace{4pt}\begin{tabular}[t]{@{\extracolsep{\fill}}ll@{}}
\multicolumn{2}{c}{H$_3$C$-$CH$_3$\vphantom{\LARGE{A}}}\\[3pt]
$r$/\AA & $V_{\sigma}(r)/{\rm eV}$\\
\hline\\[-9pt]
1.0000 &      11.3730 \\
1.0500   &    8.31871\\
\mbox{}& \\
\mbox{}& \\
1.2000  &   2.84798\\
\mbox{}& \\
1.2500    &  1.85257\\
1.3000     &  1.13216\\
\mbox{}& \\
1.3500      & 0.62916\\
1.4000      & 0.29792\\
1.4500      & 0.10209\\
1.5000      & 0.01258\\
1.5290 & 0\\
1.5500       & 0.00609\\
1.6000       & 0.06391\\
1.6500       & 0.17101\\
1.7000       & 0.31534\\
1.7500      & 0.48722\\
1.8000      & 0.67891\\
1.8500      & 0.88423\\
1.9000      & 1.09820\\
1.9500      & 1.31686\\
2.0000     & 1.53703
\end{tabular}
\hspace{2pt}&\hspace{2pt}
\begin{tabular}[t]{@{}ll@{}}
\multicolumn{2}{c}{H$_2$C$=$CH$_2$}\\[3pt]
$r$/\AA & $V_{\sigma\pi}(r)/{\rm eV}$\\
\hline\\[-9pt]
1.0000 &       6.68729\\
1.0500 &       4.33523\\
1.1000 &       2.65165\\
1.1500 &       1.48570\\
1.2000 &       0.72002\\
\mbox{}& \\
1.2500 &       0.26309\\
1.3000 &       0.04336\\
1.3372 & 0\\
1.3500 &       0.00474\\
1.4000 &       0.10310\\
1.4500 &       0.30366\\
1.5000 &       0.57893\\
\mbox{}& \\
1.5500 &       0.90718\\
1.6000 &       1.27127
\end{tabular}
\hspace{2pt}&\hspace{2pt}
\begin{tabular}[t]{@{}ll@{}}
\multicolumn{2}{c}{HC$\equiv$CH}\\[3pt]
$r$/\AA & $V_{\sigma\pi\pi}(r)/{\rm eV}$\\
\hline\\[-9pt]
1.0000 &         3.42561\\
1.0500 &         1.78388\\
1.1000 &         0.75802\\
1.1500 &         0.20275\\
1.2000 &         0.00485\\
1.2097 &  0\\
1.2500 &         0.07594\\
1.3000 &         0.34691\\
\mbox{}& \\
1.3500 &         0.76358\\
1.4000 &         1.28340\\
1.4500 &         1.87290
\end{tabular}\\
\hline
\end{tabular}
\end{tabular}
\caption{Ground-state energy 
of ethane (H$_3$C$-$CH$_3$), $V_{\sigma}(r)$, 
ethene (ethylene, H$_2$C$=$CH$_2$), $V_{\sigma\pi}(r)$,
and ethyne (acetylene, HC$\equiv$CH), $V_{\sigma\pi\pi}(r)$, in units of eV
as a function of the carbon-carbon distance~$r$ in units of~\AA.
The energies are shifted by a constant so that the potential is zero
at the equilibrium distance.
(a) The whole molecule is relaxed;
(b) The carbon-hydrogen bonds are fixed
at their positions for the equilibrium structure.\label{tab:qcdata}}
\end{table*}

Apparently, H\"uckel-PPP models can also be applied to small molecules
with $\pi$~bonds so that the applicability and accuracy of the model description
can be tested against results from quantum chemistry.
In this work, we use the two-site 
Hubbard model to describe the $\pi$-bonds
in ethene (H$_2$C$=$CH$_2$), and ethyne (HC$\equiv$CH).
We find that the H\"uckel approach provides
a reasonably accurate description of the $\pi$-bonds.
The application to trans-PA shows, however, that some adjustments
in the $\sigma$-bond spring constant 
are necessary to reproduce the experimentally observed bond lengths.

Our work is organized as follows. 
In Sect.~\ref{sec:potentials} we use data from quantum chemistry calculations
to express the ground-state energies $V_{\sigma}(r)$, $V_{\sigma\pi}(r)$, and
$V_{\sigma\pi\pi}(r)$ for $\sigma$-bonds 
and $\pi$-bonds in ethane, ethene, and ethyne.
In Sect.~\ref{sec:HHOmodel} we use the H\"uckel--Hubbard-Ohno
Hamiltonian to model the energy of the $\pi$-bonds in ethene and ethyne.
We shall see that this is possible with an accuracy of about 5\%.
However, the Hubbard-Ohno parameters for the Coulomb interaction 
remain essentially undetermined.
In Sect.~\ref{sec:PAapplication}, to test and adjust our parameter set, 
we determine the bond lengths and 
the energies of elementary electronic excitations in trans-PA.
Short conclusions, Sect.~\ref{sec:conclusions}, close our presentation.

\section{Energy of carbon single, double, and triple bonds}
\label{sec:potentials}

\subsection{Data from quantum chemistry}

We employ the quantum chemistry code 
{\sc Molpro}~\cite{MolPro} using the CCSD(T) method 
with the cc-pVTZ basis.~\cite{basis}
We calculate the energy of the molecules ethane (H$_3$C$-$CH$_3$),
ethene (a.k.a.\ ethylene, H$_2$C$=$CH$_2$), and ethyne (a.k.a.\ acetylene,
HC$\equiv$CH) as a function of the carbon-carbon distance~$r$.
The data for the energies are listed in Table~\ref{tab:qcdata}.
Similar calculations have
been done recently for ethyne,~\cite{Knowles}
and for C$_2$H$_{2n}$ for $n=0,1,2,3$.~\cite{Frenkingdata,LegezaMottet}
We can (a)~choose to relax the whole molecule,
or (b)~to fix the carbon-hydrogen bonds 
at their positions for the equilibrium structure.
The data from quantum chemistry in Table~\ref{tab:qcdata}
show that the changes are marginal for
ethyne and still very small for ethene. This could be expected because
the carbon-hydrogen bonds in the corresponding linear/planar structures
do not interfere when we stretch or shrink the carbon-carbon bond.
For ethane with its three-dimensional structure, however, the
discrepancies in energy are noticeable. 
Since we are interested in the properties of the $\sigma$-bond, we later
use the data for fixed carbon-hydrogen bonds and
plot the corresponding energies from Table~\ref{tab:qcdata}(b)
as a function of distance in Fig.~\ref{fig:potentials}.

\begin{table*}[t]
\begin{tabular}{@{}cc@{}}
(a) & (b) \\[3pt]
\begin{tabular}{|@{\hspace{3pt}}c|@{\hspace{3pt}}c|@{\hspace{3pt}}c|}
\hline
\begin{tabular}[t]{@{}lr@{}}
\multicolumn{2}{c}{\vphantom{\LARGE A}H$_3$C$-$CH$_3$}\\[3pt]
$D_l^{\sigma}$ & [eV/\AA$^{l+1}$] \\
\hline\\[-9pt]
\vphantom{\LARGE A}$D_1^{\sigma} $& 26.3496 \\
$D_2^{\sigma} $&  $-$69.9253 \\
$D_3^{\sigma} $&104.1792  \\
$D_4^{\sigma} $&$-$119.1238  \\
$D_5^{\sigma} $& 105.2503\\
$D_6^{\sigma} $& $-$70.7506 \\
$D_7^{\sigma} $&  135.5519 \\
$D_8^{\sigma} $& $-$146.0841 \\
$D_9^{\sigma} $&$-$86.5683
\end{tabular}
&
\begin{tabular}[t]{@{}lr@{}}
\multicolumn{2}{c}{\vphantom{\LARGE A}H$_2$C$=$CH$_2$}\\[3pt]
$D_l^{\sigma\pi}$ & [eV/\AA$^{l+1}$] \\
\hline\\[-9pt]
\vphantom{\LARGE A}$D_1^{\sigma\pi} $&57.4115   \\
$D_2^{\sigma\pi} $& $-$162.4340 \\
$D_3^{\sigma\pi} $& 259.2015 \\
$D_4^{\sigma\pi} $& $-$336.1312 \\
$D_5^{\sigma\pi} $&  284.6613\\
$D_6^{\sigma\pi} $& $-$116.8712\\
$D_7^{\sigma\pi} $&  816.7895
\end{tabular}
&
\begin{tabular}[t]{@{}lr@{}}
\multicolumn{2}{c}{\vphantom{\LARGE A}HC$\equiv$CH}\\[3pt]
$D_l^{\sigma\pi\pi}$ & [eV/\AA$^{l+1}$] \\
\hline\\[-9pt]
\vphantom{\LARGE A}$D_1^{\sigma\pi\pi} $&101.0452 \\
$D_2^{\sigma\pi\pi} $& $-$294.7982\\
$D_3^{\sigma\pi\pi} $&  482.5564\\
$D_4^{\sigma\pi\pi} $&  $-$ 656.0314\\
$D_5^{\sigma\pi\pi} $& 664.8695
\end{tabular}\\
\hline
\end{tabular}
&
\begin{tabular}{|@{\hspace{3pt}}c|@{\hspace{3pt}}c|@{\hspace{3pt}}c|}
\hline
\begin{tabular}[t]{@{}lr@{}}
\multicolumn{2}{c}{\vphantom{\LARGE A}H$_3$C$-$CH$_3$}\\[3pt]
$D_l^{\sigma}$ & [eV/\AA$^{l+1}$] \\
\hline\\[-9pt]
\vphantom{\LARGE A}$D_1^{\sigma} $& 28.5307 \\
$D_2^{\sigma} $&  $-$72.8851 \\
$D_3^{\sigma} $&106.8931 \\
$D_4^{\sigma} $&$-$122.0097  \\
$D_5^{\sigma} $& 111.0721\\
$D_6^{\sigma} $& $-$81.3086 \\
$D_7^{\sigma} $& 131.5314 \\
$D_8^{\sigma} $& $-$136.7223 \\
$D_9^{\sigma} $&$ -$63.3005
\end{tabular}
&
\begin{tabular}[t]{@{}lr@{}}
\multicolumn{2}{c}{\vphantom{\LARGE A}H$_2$C$=$CH$_2$}\\[3pt]
$D_l^{\sigma\pi}$ & [eV/\AA$^{l+1}$] \\
\hline\\[-9pt]
\vphantom{\LARGE A}$D_1^{\sigma\pi} $& 58.5984  \\
$D_2^{\sigma\pi} $& $-$164.9132\\
$D_3^{\sigma\pi} $&  263.5307\\
$D_4^{\sigma\pi} $& $-$342.9333 \\
$D_5^{\sigma\pi} $&  296.9368  \\
$D_6^{\sigma\pi} $& $-$141.5545 \\
$D_7^{\sigma\pi} $& 815.7742
\end{tabular}
&
\begin{tabular}[t]{@{}lr@{}}
\multicolumn{2}{c}{\vphantom{\LARGE A}HC$\equiv$CH}\\[3pt]
$D_l^{\sigma\pi\pi}$ & [eV/\AA$^{l+1}$] \\
\hline\\[-9pt]
\vphantom{\LARGE A}$D_1^{\sigma\pi\pi} $& 101.0741  \\
$D_2^{\sigma\pi\pi} $& $-$294.8085 \\
$D_3^{\sigma\pi\pi} $&   482.7707 \\
$D_4^{\sigma\pi\pi} $&  $-$656.2767 \\
$D_5^{\sigma\pi\pi} $&  664.6464
\end{tabular}\\
\hline
\end{tabular}
\end{tabular}
\caption{Parameters $D_l$ ($l\leq 9$) for the force fields $F_{\sigma}(r)$,
$F_{\sigma\pi}(r)$, and $F_{\sigma\pi\pi}(r)$,
of ethane, ethene, and ethyne, respectively. 
(a) The whole molecule is relaxed;
(b) The carbon-hydrogen bonds are fixed
at their positions for the equilibrium structure.\label{tab:Kvalues}}
\end{table*}

\subsection{Polynomial parameterization}

First, we compare the bond lengths with the experimental data.
We find for ethane (one $\sigma$-bond), ethene (one $\sigma$-bond
plus one $\pi$-bond), and ethyne (one $\sigma$-bond
plus two $\pi$-bonds),
\begin{equation}
\begin{array}{@{}rcl@{}}
r_{\sigma}&=& 1.5290\, \hbox{\AA} \; ,\\
r_{\sigma\pi}&=& 1.3372\, \hbox{\AA} \;,\\
r_{\sigma\pi\pi}&=& 1.2097\, \hbox{\AA} \;,
\end{array}
\quad 
\begin{array}{@{}rcl@{}}
r_{\sigma}^{\rm exp}&=& 1.5360\, \hbox{\AA}
\quad \hbox{[\onlinecite{Herzberg}]}\, ,\\
r_{\sigma\pi}^{\rm exp}&=& 1.3390\, \hbox{\AA} 
\quad \hbox{[\onlinecite{Herzberg}]}\, ,\\
r_{\sigma\pi\pi}^{\rm exp}&=& 1.203\hphantom{0}\, \hbox{\AA} 
\quad \hbox{[\onlinecite{Lievin}]}\, ,
\end{array}
\end{equation}
in very good agreement with the experimental values,
with deviations of less than~0.5\%.

Next, we express the energies $V_{\rm e}(r)$ as a Taylor series
around the equilibrium distances~$r_{\rm e}$,
\begin{equation}
V_{\rm e}(r)=V_{0}^{\rm e}+\sum_{n=2}^{10}\frac{D^{\rm e}_{n-1}}{n}(r-r_{\rm e})^n \; ,
\label{eq:defVsigmaetc}
\end{equation}
where ${\rm e}=\sigma,\sigma\pi, \sigma\pi\pi$ for the three molecules,
and $V_{0}^{\rm e}$ is independent of~$r$.
In the following, we focus on the force fields $F_{\rm e}(r)=-V_{\rm e}'(r)$,
\begin{equation}
F_{\rm e}(r)=-\sum_{n=1}^{9}D_n^{\rm e}(r-r_{\rm e})^n \; ,
\end{equation}
because they enter the optimization equations.
The parameter sets for ethane, ethene, and ethyne are collected 
in Table~\ref{tab:Kvalues}.

One of the vibrational modes in ethane, ethene, and ethyne
with reciprocal wave length (wave number) $1/\lambda^{\rm e}$ 
can be assigned to a bare carbon-carbon stretch mode.
They can be compared to the values from quantum-chemistry
calculations,
\begin{equation}
\begin{array}{@{}rcl@{}}
1/\lambda_{\sigma}&=&1013\, {\rm cm}^{-1} \, ,\\
1/\lambda_{\sigma\pi}&=& 1671\, {\rm cm}^{-1}\, ,\\
1/\lambda_{\sigma\pi\pi}&=& 2000\, {\rm cm}^{-1}\, ,
\end{array}
\;\;
\begin{array}{@{}rcl@{}}
1/\lambda^{\rm exp}_{\sigma}&=&992.9\, {\rm cm}^{-1} 
\quad \hbox{[\onlinecite{Helvoort}]}\, ,\\
1/\lambda^{\rm exp}_{\sigma\pi}&=& 1623\hphantom{.}\, {\rm cm}^{-1}
\quad \hbox{[\onlinecite{Krasser}]}\, ,\\
1/\lambda^{\rm exp}_{\sigma\pi\pi}&=& 1974\hphantom{.}\, {\rm cm}^{-1}
\quad \hbox{[\onlinecite{FastWelsh}]} \, .
\end{array}
\label{eq:thelambdas}
\end{equation}
The experimental values in eq.~(\ref{eq:thelambdas})
are in very good agreement 
with the results from quantum chemistry, with deviations of less than~3\%.
This is the desired 
accuracy of the fit to a H\"uckel--Hubbard-Ohno model.

The normal-mode frequencies given in eq.~(\ref{eq:thelambdas})
correspond to vibrational modes that involve the movement of all atoms
in the molecule. Instead, one may want to invoke
a simpler oscillator model with fixed
carbon-hydrogen bonds, a single force constant~$\tilde{D}_1^{\rm e}$,
and an effective mass~$m^{\rm e}=(12+n)u/2$ (atomic mass unit:
$u=1.66054\cdot 10^{-27}\, {\rm kg}$; $n=1,2,3$ 
for ethane, ethene, ethyne).~\cite{largeKinPA}
Then, the force constant obeys
\begin{equation}
\tilde{D}_1^{\rm e}=m^{\rm e}c^2\left(
\frac{2\pi}{\lambda_{{\rm CC}}^{\rm e}}\right)^2 \;.
\end{equation}
A fit to the experimental data for ethane, ethene, and ethyne 
then leads to 
\begin{equation}
\begin{array}{@{}rcl@{}}
\tilde{D}_1^{\sigma}&=&27.2\, \hbox{eV/\AA$^2$} \, ,\\
\tilde{D}_1^{\sigma\pi}&=&67.8\, \hbox{eV/\AA$^2$} \, ,\\
\tilde{D}_1^{\sigma\pi\pi}&=&93.1\, \hbox{eV/\AA$^2$} \, ,
\end{array}
\;\;
\begin{array}{@{}rcl@{}}
D_1^{\sigma}&=&\hphantom{0}26.3\, \hbox{eV/\AA$^2$}  \, , \\
D_1^{\sigma\pi}&=&\hphantom{0}57.4\, \hbox{eV/\AA$^2$} \, , \\
D_1^{\sigma\pi\pi}&=&101.0\, \hbox{eV/\AA$^2$} \,, 
\end{array}
\label{eq:theDs}
\end{equation}
where we used $1\, {\rm N/m}=0.01\, {\rm mdyn}/\hbox{\AA}=
1/16.022\, {\rm eV}/\hbox{\AA}{}^2$ for the conversion of units.
The comparison shows that oscillator models with a single `spring constant' 
shall employ values $\tilde{D}_1$ that deviate from
the bare values $D_1$ from quantum chemistry calculations by 10\%--15\%.

\section{H\"uckel--Hubbard-Ohno model}
\label{sec:HHOmodel}

\subsection{Hamiltonian for a $\boldsymbol{\pi}$-bond}

We model a $\pi$-bond
by a two-site single-orbital H\"uckel--Hubbard-Ohno model,
\begin{eqnarray}
\hat{H}_{\pi}(r)&=&
-t(r) 
\left(
\hat{c}_{1,\uparrow}^{\dagger}\hat{c}_{2,\uparrow}^{\vphantom{\dagger}}
+ \hat{c}_{2,\uparrow}^{\dagger}\hat{c}_{1,\uparrow}^{\vphantom{\dagger}}
+
\hat{c}_{1,\downarrow}^{\dagger}\hat{c}_{2,\downarrow}^{\vphantom{\dagger}}
+ \hat{c}_{2,\downarrow}^{\dagger}\hat{c}_{1,\downarrow}^{\vphantom{\dagger}}
\right)
\nonumber \\
&& +U\left( \hat{n}_{1,\uparrow}\hat{n}_{1,\downarrow}
+ \hat{n}_{2,\uparrow}\hat{n}_{2,\downarrow}\right)
\nonumber \\
&& 
+ V(r) \left( \hat{n}_1-1\right)\left( \hat{n}_2-1\right) 
\; ,
\end{eqnarray}
where $\hat{c}_{l,\sigma}^{\dagger}$ ($\hat{c}_{l,\sigma}^{\vphantom{\dagger}}$)
creates (annihilates) an electron with spin $\sigma=\{\uparrow,\downarrow\}$
on carbon atom $l=1,2$. Moreover, 
$\hat{n}_{l,\sigma}=\hat{c}_{l,\sigma}^{\dagger}
\hat{c}_{l,\sigma}^{\vphantom{\dagger}}$ counts the number of $\sigma$-electrons
on carbon atom~$l$ and $\hat{n}_l=\hat{n}_{l,\uparrow}+\hat{n}_{l,\downarrow}$.
We assume that the $\sigma$-bond is rigid and
contributes the potential $V_{\sigma}(r)$ to the energy.

The Hamiltonian parametrically depends on the distance~$r$ between
the carbon atoms. The electron transfer parameter decreases
exponentially as a function of distance,
\begin{equation}
t_{\alpha}(r) = t_0 \exp \left( -\alpha (r-r_0)/t_0 \right)\; ,
\label{eq:Peierlstofr}
\end{equation}
with $t_ {\alpha}(r_0)
=t_0\equiv 2.5\, {\rm eV}$ at distance $r=r_0\equiv1.4\, \hbox{\AA}$. 
The strength of the Peierls coupling~$\alpha$
and the values for the Coulomb interaction $U$ and~$V$
are fit parameters.
The distance-dependence of the density-density interaction 
is given by the Ohno expression~\cite{Baeriswyl,Barford}
\begin{equation}
V(r)= \frac{V}{\sqrt{1+\beta(r/\hbox{\AA})^2}} \; ,
\beta= \left(\frac{V}{14.397\, {\rm eV}}\right)^2 \; .
\label{eq:Ohnoexpression}
\end{equation}
The Ohno form guarantees that, at large distances, 
the electrons interact via their unscreened
Coulomb interaction, $e^2=14.397\, {\rm eV}\hbox{\AA}$.

\begin{figure}[b]
\begin{center}
\mbox{}\\[9pt]
\hspace*{-36pt}
\includegraphics[width=7.5cm]{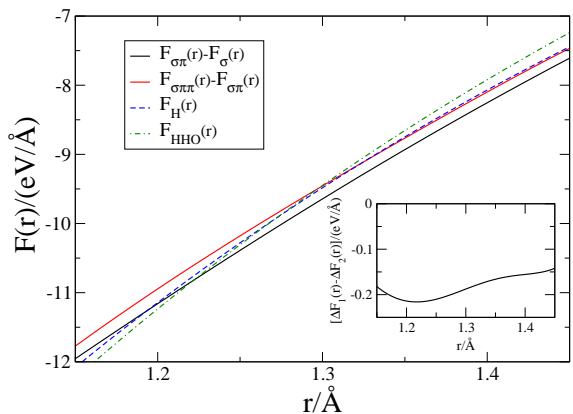}
\end{center}
\caption{(Color online) 
Force function differences $\Delta F_1(r)=F_{\sigma\pi}(r)-F_{\sigma}(r)$,
$\Delta F_2(r)=F_{\sigma\pi\pi}(r)-F_{\sigma\pi}(r)$
due to the $\pi$-bonds in ethene and in ethyne
for carbon distances $1.15\,\hbox{\AA}<r<1.45\, \hbox{\AA}$.
The force function differences are not the same,
they deviate from each other by about $0.2\, {\rm eV}$ 
for $1.15\,\hbox{\AA}<r<1.45\, \hbox{\AA}$,
as seen in the inset.
The dashed lines show the fit from the H\"uckel model, 
$F_{\rm H}(r)=-E'_{\pi}(r,4.035\, {\rm eV}/\hbox{\AA},0,0)$
and from the H\"uckel--Hubbard-Ohno model,
$F_{\rm HHO}(r)=
-E'_{\pi}(r,4.119\, {\rm eV}/\hbox{\AA},6\, {\rm eV},3\, {\rm eV})$.
\label{fig:forcefields}}
\end{figure}

The ground-state energy of the H\"uckel--Hubbard-Ohno model is denoted
by $E_{\pi}(r,\alpha,U,V)$. 
The two-site model is readily diagonalized,~\cite{BarfordPDA} 
\begin{equation}
E_{\pi}(r,\alpha,U,V)=
\frac{U-V(r)}{2} -\sqrt{\frac{(U-V(r))^2}{4}+4 [t_{\alpha}(r)]^2} \; .
\label{eq:Esigmapi}
\end{equation}
Up to a constant energy shift,
$V_{\sigma}(r)+E_{\pi}(r,\alpha,U,V)$ defines the H\"uckel--Hubbard-Ohno
approximation to the energy function for ethene, $V_{\sigma\pi}(r)$.

For ethyne, we further assume that the two $\pi$-bonds are independent
of each other. An explicit calculation
of the ground-state energy of a two-site model with four orbitals per site
shows that, for moderate Coulomb interactions, the ground-state energy
can indeed be approximated as the sum of two independent
two-orbital models. Therefore, the H\"uckel--Hubbard-Ohno approach
predicts
\begin{equation}
V_{\sigma\pi}(r)-V_{\sigma}(r) +C_1 
= V_{\sigma\pi\pi}(r)-V_{\sigma\pi}(r) +C_2 \; .
\label{eq:testhueckel}
\end{equation}
The quality of this hypothesis can be tested by comparing
the corresponding differences in the force functions,
$\Delta F_1(r)=F_{\sigma\pi}(r)-F_{\sigma}(r)$ and 
$\Delta F_2(r)=F_{\sigma\pi\pi}(r)-F_{\sigma\pi}(r)$.

In Fig.~\ref{fig:forcefields} we show 
the two force function differences $\Delta F_i(r)$
for carbon-carbon distances $1.15\, \hbox{\AA}<r<1.45\, \hbox{\AA}$.
As is seen from the inset, they differ
by about $0.2\,{\rm eV}/\hbox{\AA}$.
Similarly, the H\"uckel force field $F_{\rm H}(r)$ at $\alpha=4.035\, {\rm eV}$
and the H\"uckel--Hubbard-Ohno force field 
$F_{\rm HHO}(r)$ at $\alpha=4.119\, {\rm eV}$
and $U=2V=6\, {\rm eV}$ deviate from $\Delta F_i(r)$ by several tenth
of an ${\rm eV}/\hbox{\AA}$.
Since $\Delta F_i(r)$ is of the order of $10\, {\rm eV}/\hbox{\AA}$,
the H\"uckel--Hubbard-Ohno description of the 
$\pi$-bonds in ethene and ethyne is accurate within a few percent
which is of the same order of magnitude as the 
accuracy of the quantum chemistry data for the
vibrational frequencies.
Therefore, the H\"uckel--Hubbard-Ohno description 
is accurate enough to match the quality of the quantum chemistry data.
For a further improvement, one may refine the expression~(\ref{eq:Peierlstofr})
for the distance-dependence of the electron transfer parameter.~\cite{Lepetit}

\subsection{Parameter optimization}
\label{subsec:optimizeparameters}

Since the force functions $\Delta F_i(r)$ differ, we cannot determine
$E_{\pi}(r,\alpha,U,V)$ from eq.~(\ref{eq:testhueckel}) as the 
difference between $V_{\sigma}(r)$ and $V_{\sigma\pi}(r)$.
Instead, we have to set up an optimization scheme.
We define the cost function (all lengths in~\AA, all energies in~eV)
\begin{eqnarray}
W(\alpha,U,V)&=&
\int_{1.15}^{1.45}{\rm d}r 
\left(F_{\sigma\pi}(r)-F_{\sigma}(r) +E_{\pi}'(r)\right)^2 \nonumber \\
&& +\int_{1.15}^{1.45}{\rm d}r \left(F_{\sigma\pi\pi}(r)-F_{\sigma\pi}(r)
+E_{\pi}'(r)\right)^2  ,\nonumber \\
\end{eqnarray}
where $E_{\pi}'(r)=({\rm d}E_{\pi}(r,\alpha,U,V))/({\rm d} r)$
is the negative force field due to the $\pi$-bond 
in the H\"uckel--Hubbard-Ohno description.
Using this cost function for the parameter optimization, 
the H\"uckel--Hubbard-Ohno model will provide a suitable
description of the $\pi$-bonds in ethene and ethyne.

As a first step, we seek the optimal value for the bare H\"uckel model, i.e.,
we optimize $W(\alpha,0,0)$. The result is 
$\alpha^{\rm H}=4.035\, {\rm eV}/\hbox{\AA}$.
The H\"uckel model with $\alpha^{\rm H}$
provides a good description of the $\pi$-bonds, see Fig.~\ref{fig:forcefields}.
The bare H\"uckel force field, 
$F_{\rm H}(r)=-E_{\pi}'(r,4.035\, {\rm eV}/\hbox{\AA},0,0)$
deviates from the force fields $\Delta F_{1,2}(r)$ 
by only a few tenth of an eV/\AA.
As a consequence, the H\"uckel model reproduces the bond lengths~$r_{\sigma\pi}$
and $r_{\sigma\pi\pi}$
and the parameters $D_1^{\sigma\pi}$ and $D_1^{\sigma\pi\pi}$
from quantum chemistry calculations with good accuracy, 
see Table~\ref{tab:HHOfit}.

The success of the bare H\"uckel model 
indicates that the Hubbard-Ohno interaction cannot improve the results very much.
More importantly, 
for moderate values for $V$ ($V<6\, {\rm eV}$) 
and  for $r$ of the order of one Angstr\o m, the dependence of $V(r)$
on~$r$ is fairly small, and
the range of acceptable values for $U$ and $V$ is quite large
for the optimization functions chosen.
Correspondingly, short-range parameterizations of the PPP potential
can give parameter values very different from ours.~\cite{Chinesenfit}
Given the finite accuracy of the quantum chemistry data,
it is therefore difficult to derive reliable values for~$U$ and~$V$
from our fits.

\begin{table}[t]
\begin{center}
\begin{tabular}{@{}ccccc@{}}
& \begin{tabular}{@{}c@{}}
$r_{\sigma\pi}$\\
\hline\\[-8pt]
\AA\end{tabular}
 & \begin{tabular}{@{}c@{}}
$D_1^{\sigma\pi}$\\
\hline\\[-8pt]
eV/\AA$^2$\end{tabular} 
& \begin{tabular}{@{}c@{}}
$r_{\sigma\pi\pi} $\\
\hline\\[-8pt]
\AA
\end{tabular} & \begin{tabular}{@{}c@{}}
$D_1^{\sigma\pi\pi}$\\
\hline\\[-8pt]
eV/\AA$^2$\end{tabular} \\
\hline\\[-9pt]
Experiment & 1.3390 &  & 1.203\hphantom{0} &  \\
Quantum Chemistry & 1.3372 & 58.60 & 1.2097 & 101.07\\
H\"uckel model ($U=V=0$) & 1.3403 & 57.28 &1.2079  & 100.10 \\
H\"uckel--Hubbard-Ohno model & 1.3417 & 55.66 & 1.2071 & \hphantom{0}99.28
\end{tabular}
\end{center}
\caption{Equilibrium distances and force parameters for ethene and ethyne
from experiment, quantum chemistry CCSD(T)/cc-pVTZ, the H\"uckel model
 with $\alpha=4.035\, {\rm eV}/\hbox{\AA}$,
and the H\"uckel--Hubbard-Ohno model with
$\alpha=4.119\, {\rm eV}/\hbox{\AA}$, $U=6\, {\rm eV}$, 
$V=3\, {\rm eV}$.\label{tab:HHOfit}}
\end{table}

We note, however, that $U$ and~$V$ strongly influence
the size of the single-particle gap and the exciton binding energy
in polymers. Therefore, we argue that the Coulomb parameters for short molecules
should not differ much from the parameter set
$(U=6\, {\rm eV},V=3\, {\rm eV})$ as obtained from our analysis of the PDA
spectra.~\cite{BarfordPDA}
For this reason, we restrict ourselves to the optimization of 
$W(\alpha,6\, {\rm eV},3\, {\rm eV})$,
and obtain $\alpha^{\rm HHO}= 4.119\, {\rm eV}/\hbox{\AA}$ as the optimal
value. As for the bare H\"uckel model,
the H\"uckel--Hubbard-Ohno model reproduces the bond lengths
and force parameters  from quantum chemistry
calculations with good accuracy, see Table~\ref{tab:HHOfit}.
The overall agreement with the force functions
is also quite good for all
carbon-carbon distances $1.15\, \hbox{\AA}<r<1.45\, \hbox{\AA}$,
see Fig.~\ref{fig:forcefields}.

\section{Application to polyacetylene}
\label{sec:PAapplication}

\subsection{Parameters for trans-polyacetylene}

The analysis of the bond energies in our di-carbon molecules
permits two conclusions. First,
the Peierls coupling should be 
$\alpha\approx 4\, {\rm eV}/\hbox{\AA}$
which is somewhat smaller than the value proposed
by Su, Schrieffer and Heeger,
$\alpha^{\rm SSH}=4.7\, {\rm eV}/\hbox{\AA}$.~\cite{Baeriswyl,SSH}
Similar values, with a deviation of $\pm 20\%$,
were generally used for the description of $\pi$-conjugated
polymers.~\cite{Barford,smallKinPA,largeKinPA}

Second, we can expand the $\sigma$-bond potential around
the average bond distance in trans-PA
to determine the spring constant~$K_{\sigma}^{\rm PA}$.
We Taylor expand the $\sigma$-bond force field $F_{\sigma}(r)$
around $r=r_0$,
\begin{equation}
F_{\sigma}(r)=-\sum_{n=1}^{9}D_n^{\sigma}(r-r_{\sigma})^n=
-\sum_{n=0}^{9} K_{\sigma,n}(r-r_0)^n \; ,
\end{equation}
and find ($K_{\sigma,n}$ is in units of ${\rm eV}/\hbox{\AA}{}^{n+1}$) 
\begin{eqnarray}
K_{\sigma,0}=-5.16199\, , \, &K_{\sigma,1}=\hphantom{-}53.8998& \,  , \, 
K_{\sigma,2}=-129.282 \, ,\nonumber \\
K_{\sigma,3}=\hphantom{-}193.365 \, , \, &K_{\sigma,4}=-226.216& \,  , \, 
K_{\sigma,5}=\hphantom{-}234.224  \, ,\nonumber \\
K_{\sigma,6}=-252.402 \, , \, &K_{\sigma,7}=\hphantom{-}234.716& \,  , \, 
K_{\sigma,8}= -63.2203\, ,\nonumber \\
\end{eqnarray}
and $K_{\sigma,9}= -63.3005\, {\rm eV}/\hbox{\AA}{}^{10}=D_9^{\sigma}$, 
see Table~\ref{tab:Kvalues}(b).
The first term, $-K_{\sigma,0}=5.2\, {\rm eV}/\hbox{\AA}>0$,
describes the repulsive force of the $\sigma$-bond.
It opposes the shrinking of the $\sigma$-bonds in trans-PA
from $r_{\sigma}=1.529\, \hbox{\AA}$ down to $r_0=1.4\, \hbox{\AA}$.
This repulsive force must be compensated by the binding due to the
itinerant $\pi$-electrons.

The second term describes the enhanced spring constant due to the 
compressed $\sigma$-bond,
$K_{\sigma,1}=54\, {\rm eV}/\hbox{\AA}{}^2$, so that
we should set
$K_{\sigma}^{\rm PA}=K_{\sigma,1}$.
This value can be compared 
with the literature values
used for trans-polyacetylene.~\cite{smallKinPA,largeKinPA}
Our analysis of the $\sigma$-bond potential in ethane, $V_{\sigma}(r)$,
supports the larger value derived in Ref.~[\onlinecite{largeKinPA}],
$K_{\sigma}^{{\rm PA},2}=46\, {\rm eV}/\hbox{\AA}{}^2$.
Moreover, we find $\alpha/K_{\sigma}^{\rm PA}\approx 0.08\, \hbox{\AA}$ which
implies that the electron-phonon coupling constant,
\begin{equation}
\lambda=\frac{2\alpha^2}{\pi t_0K_{\sigma}^{\rm PA}} \; ,
\end{equation}
is small, $\lambda\approx 0.08$.~\cite{Baeriswyl} 
Therefore, polaronic effects are 
predicted to be of minor importance for trans-polyacetylene,
in contrast to the SSH picture, $\lambda^{\rm SHH}=0.2$.~\cite{SSH}

\subsection{H\"uckel model}

The values for the electron-phonon coupling and the $\sigma$-bond 
potential $V_{\sigma}(r)$ are important for the theoretical description
of the bond lengths in~PA. We find that the simple H\"uckel
model cannot account for the dimerization in trans-polyacetylene.

\subsubsection{Hamiltonian}

We consider $L$ unit cells with two carbon atoms each.
The spin-1/2 electrons move on sites~$l=1,2,\ldots,2L$, 
and periodic boundary conditions apply. 
The system is half filled, i.e., the number of electrons equals the number of sites,
$N_{\uparrow}+N_{\downarrow}=2L$; the system is paramagnetic, 
$N_{\uparrow}=N_{\downarrow}=L$.
In the H\"uckel description, the electrons move between neighboring sites,
\begin{eqnarray}
\hat{T}&=& -\sum_{\sigma}\sum_{l=1}^L \Bigl[
t_{\rm odd}
\left( 
\hat{c}_{2l-1,\sigma}^{\dagger}\hat{c}_{2l,\sigma}^{\vphantom{+}} 
+
\hat{c}_{2l,\sigma}^{\dagger}\hat{c}_{2l-1,\sigma}^{\vphantom{+}} 
\right) \nonumber \\
&& \hphantom{ -\sum_{\sigma}\sum_{l=1}^L }+
t_{\rm even}
\left( 
\hat{c}_{2l,\sigma}^{\dagger}\hat{c}_{2l+1,\sigma}^{\vphantom{+}} 
+
\hat{c}_{2l+1,\sigma}^{\dagger}\hat{c}_{2l,\sigma}^{\vphantom{+}} 
\right)
\Bigr] \; ,\nonumber \\
\end{eqnarray}
where
\begin{equation}
t_{\rm odd}=t_{\alpha}(r_{\rm s}) \quad, \quad
t_{\rm even}=t_{\alpha}(r_{\rm d})
\end{equation}
are the electron transfer matrix elements for the single-bonds and double-bonds,
respectively. In this Ansatz we take into account the bond dimerization (Peierls effect).
The lengths of the single and double bonds are given by
\begin{equation}
r_{\rm s}=r_{\sigma}-s+\Delta
\quad, \quad
r_{\rm d}=r_{\sigma}-s-\Delta \; .
\end{equation}
Here, $s$ describes the average bond-length reduction from $r_{\sigma}$
to $r_{\sigma}-s$, and $\Delta$ describes the bond-length alternation.

The total Hamiltonian is given by the sum of the electrons' kinetic energy
and the potential energy contribution from $V_{\sigma}(r)$,
\begin{eqnarray}
\hat{H}^{\rm H}(s,\Delta)&=&\hat{T}(s,\Delta) + L V_{\rm bond}(s,\Delta)  \nonumber \; ,\\
V_{\rm bond}(s,\Delta)&=&V_{\sigma}(r_{\sigma}-s+\Delta)+ 
V_{\sigma}(r_{\sigma}-s-\Delta)  .
\label{eq:Hueckelhamilt}
\end{eqnarray}
At the optimal values of $s$ and $\Delta$, 
the ground-state energy of $\hat{H}^{\rm H}$
has its minimum.
Experimentally,~\cite{NMR} we have $r_{\rm s}=1.44\, \hbox{\AA}$ and
$r_{\rm d}=1.36\, \hbox{\AA}$ 
with $(r_{\rm s}+r_{\rm d})/2=r_0=1.40\, \hbox{\AA}$,
so that
$\Delta_0^{\rm exp}=(r_{\rm s}-r_{\rm d})/2=0.04\, \hbox{\AA}$
and $s_0^{\rm exp}=r_{\sigma}-r_0=0.136\, \hbox{\AA}$.

\subsubsection{Ground-state energy}

To determine the ground-state energy as a function of $s$ and $\Delta$,
we introduce the dimensionless units
\begin{equation}
\sigma = \frac{s\alpha}{t_0} \quad , \quad 
\delta = \frac{\Delta\alpha}{t_0} \; .
\end{equation}
The bare dispersion relation~$\epsilon(k)$ and the hybridization
function~$\Delta(k)$ are given by
\begin{eqnarray}
\epsilon(k) &=& -(t_{\rm odd}+t_{\rm even})\cos(k) 
=-2\tilde{t}_0\cos(k) e^{\sigma}\cosh(\delta)
\nonumber \; , \\
\Delta(k) &=& (t_{\rm even}-t_{\rm odd})\sin(k) =2\tilde{t}_0\sin(k)
e^{\sigma}\sinh(\delta)
 \; .
\end{eqnarray}
Here, we use the abbreviation $\tilde{t}_0=t_0\exp[-(r_{\sigma}-r_0)\alpha/t_0]$.
The H\"uckel-Peierls Hamiltonian is diagonal in reciprocal space,~\cite{SSH,PhilMag}
\begin{eqnarray}
\hat{T}
&=& \sum_{|k|\leq \pi/2,\sigma} E(k) (\hat{a}_{k,\sigma,+}^{\dagger}
\hat{a}_{k,\sigma,+}^{\vphantom{\dagger}}
- \hat{a}_{k,\sigma,-}^{\dagger} \hat{a}_{k,\sigma,-}^{\vphantom{+}})
\; , \nonumber \\
E(k) &=& \sqrt{[\epsilon(k)]^2+[\Delta(k)]^2}  \nonumber \\
&=&
2\tilde{t}_0
e^{\sigma}\sqrt{[\cosh(\delta)]^2-[\sin(k)]^2}
\; .
\label{Tdiapeierls}
\end{eqnarray}
Here, $\pm E(k)$ is the dispersion relation for the upper~($+$)
and lower~($-$) Peierls band.
The ground state is the Peierls insulator,
\begin{equation}
|\Psi_0\rangle 
= \prod_{\sigma}
\prod_{|k|\leq \pi/2} \hat{a}_{k,\sigma,-}^{\dagger} | {\rm vac}\rangle \; ,
\end{equation}
where the lower Peierls band is completely filled.
Note that the wave numbers~$k$ are quantized in units of $2\pi/(2L)$ because the
chain has $2L$ sites.

The sum of the electrons' kinetic energy and of the lattice potential energy
per unit cell is given by
\begin{eqnarray}
e_{\rm tot}(\sigma,\delta)&=& T(\sigma,\delta)/L+
V_{\rm bond}(t_0\sigma/\alpha,t_0\delta/\alpha) \; , \nonumber \\
T(\sigma,\delta)/L&=& 
-4 \tilde{t}_0 e^{\sigma}
\int_{-\pi/2}^{\pi/2}\frac{{\rm d}k}{\pi}  \sqrt{[\cosh(\delta)]^2-[\sin(k)]^2} \; .
\nonumber \\
\end{eqnarray}
A factor of two in the kinetic energy accounts for the spin degeneracy.
The minimization with respect to $\sigma$ gives
\begin{eqnarray}
-\frac{T(\sigma_0,\delta_0)}{L}&=&
4 \tilde{t}_0 e^{\sigma_0}
\int_{-\pi/2}^{\pi/2}\frac{{\rm d}k}{\pi}  \sqrt{[\cosh(\delta_0)]^2-[\sin(k)]^2}
\nonumber \\
&=& \frac{t_0}{\alpha}
\Bigl[
F_{\sigma}(r_{\sigma}-(t_0/\alpha)\sigma_0+(t_0/\alpha)\delta_0)\nonumber \\
&& \phantom{\Bigl[}+ 
F_{\sigma}(r_{\sigma}-(t_0/\alpha)\sigma_0-(t_0/\alpha)\delta_0)
\Bigr]
\; .
\label{eq:firstmin}
\end{eqnarray}
The minimization with respect to $\delta$ leads to
\begin{eqnarray}
\frac{T^{\prime}(\sigma_0,\delta_0)}{L}&=&
-4 \tilde{t}_0 e^{\sigma_0}
\int_{-\pi/2}^{\pi/2} \frac{{\rm d}k}{\pi}
\frac{\cosh(\delta_0)\sinh(\delta_0)}{\sqrt{[\cosh(\delta_0)]^2-[\sin(k)]^2}}
\nonumber \\
&=&
\frac{t_0} {\alpha}
\Bigl[
F_{\sigma}(r_{\sigma}-(t_0/\alpha)\sigma_0+(t_0/\alpha)\delta_0) \nonumber \\
&& \hphantom{\frac{t_0} {\alpha}\Bigl[}-
F_{\sigma}(r_{\sigma}-(t_0/\alpha)\sigma_0-(t_0/\alpha)\delta_0)
\Bigr]
\; .
\label{eq:secondmin}
\end{eqnarray}
The equations~(\ref{eq:firstmin}) and (\ref{eq:secondmin})
must be solved simultaneously to obtain the values for the ground-state
distortions $\delta_0$ and $\sigma_0$. This is done 
using {\sc Mathematica}.~\cite{Mathematica}

When we use $\alpha^{\rm H}=4.035\, {\rm eV}/\hbox{\AA}$
and $V_{\sigma}(r)$ from eq.~(\ref{eq:defVsigmaetc}) with the parameters
from Table~\ref{tab:Kvalues}, we find 
that the $\sigma$-bond shrinks by $s_0^{\rm H}=0.128\, \hbox{\AA}$ 
so that the average bond length is $r_0^{\rm H}=1.401\, \hbox{\AA}$,
in good agreement with experiment.
However, the dimerization is $\Delta_0^{\rm H}=0.004\, \hbox{\AA}$, an order of 
magnitude smaller than in experiment.
These numbers do not change significantly when
we use the linear approximation 
$F_{\sigma}(r)\approx -[K_{\sigma,0}+K_{\sigma,1}(r-r_0)]$
for $|r-r_0|\ll r_0$.
This shows that the curvature of the force field can be ignored
when we optimize the structure
in the vicinity of the equilibrium distance, $r\approx r_0$.
For given $s_0$ and $\Delta_0$, the H\"uckel-Peierls model
requires the parameters $K_{\sigma,0}=-5.2\, {\rm eV}/\hbox{\AA}$
and $K_{\sigma,1}=34.3\, {\rm eV}/\hbox{\AA}{}^2$, 
corresponding to the much smaller spring 
constant~$K_{\sigma}^{{\rm PA},1}=31\, {\rm eV}/\hbox{\AA}{}^2$
suggested for trans-polyacetylene.~\cite{smallKinPA}

The comparison shows that additional terms must be included in the 
Hamiltonian, e.g., the Coulomb interaction between the $\pi$-electrons,
that must be responsible for a large part of the bond dimerization
in trans-polyacetylene.

\subsubsection{Single-particle gap}

Experimentally, it is very difficult to determine the single-particle gap
in trans-polyacetylene because high-quality single-crystals as for PDA cannot be
fabricated. Therefore, Franz-Keldysh oscillations of unbound single-particle 
excitations have not been detected in electro-absorption experiments
on trans-polyacetylene. 

The optical
absorption of polyacetylene films becomes significant above
$E_{\rm onset}=1.5\, {\rm eV}$ and shows a peak at $E_{\rm peak}=1.9\, {\rm eV}$;
the electro-absorption measurements display
a strong signal at $\hbar\omega=1.4\, {\rm eV}$.~\cite{Orenstein,Phillips}
Third-harmonic generation is large at $\hbar\omega_{\rm T}=0.6\, {\rm eV}$
which is evidence for an 
excitonic state at $E_{\rm exc}=1.8\, {\rm eV}$,~\cite{Fann}
as also seen in combined absorption/refraction measurements,
$E_{\rm exc}=1.7\, {\rm eV}$.~\cite{Leising}
Apparently, the linear absorption spectrum of trans-polyacetylene results from
disorder-broadening of the exciton resonance.~\cite{Kubo}

We assume that the exciton binding energy in trans-polyacetylene is of the same
order of magnitude as in other conjugated polymers such as PDA, 
$\Delta_s\approx 0.4\,{\rm eV} \ldots 0.5\, {\rm eV}$.~\cite{Liess} Then, the 
single-particle gap is estimated to be
$E_{\rm gap}\approx 2.2\,{\rm eV}\ldots 2.3\, {\rm eV}$.
These estimates are supported by calculations of Mott-Wannier excitons
for correlated electrons in one dimension.~\cite{Rohlfing}

We compare these numbers with the predictions from the H\"uckel-Peierls model.
The Peierls insulator has the single-particle gap
\begin{eqnarray}
E_{\rm gap}^{\rm HP}&=&2\Delta(\pi/2)=4\tilde{t}_0e^{\sigma_0}\sinh(\delta_0)
\nonumber \\
&=&4t_0\exp[-(r_{\sigma}-r_0-s_0)\alpha/t_0]\sinh(\delta_0)\nonumber \\
&\approx& 4\alpha\Delta_0 \; .
\label{eq:gapHueckel}
\end{eqnarray}
The approximation holds because we have $r_{\sigma}-r_0\approx s_0$
and $\delta_0=\Delta_0\alpha/t_0\ll 1$. From eq.~(\ref{eq:gapHueckel}) 
we see that the combination of $\alpha^{\rm H}\approx 4\, {\rm eV}/\hbox{\AA}$
with the experimental value $\Delta_0=0.04\, \hbox{\AA}$
leads to a H\"uckel-Peierls gap of $E_{\rm gap}^{\rm HP}\approx 0.64\, {\rm eV}$.
Apparently, the Peierls gap alone cannot account for the observed single-particle 
gap in trans-polyacetylene. 

\subsection{H\"uckel--Hubbard-Ohno model}

\subsubsection{Hamiltonian}

\begin{figure}[ht]
\begin{center}
\includegraphics[height=3.3cm]{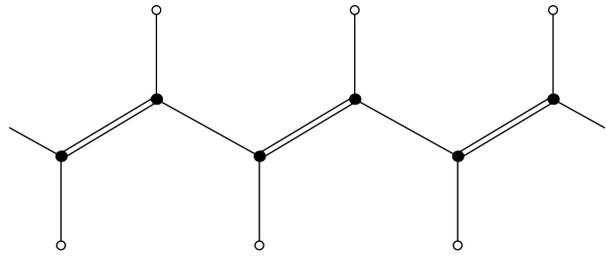}\\
\end{center}
\caption{Planar and unflexed trans-polyacetylene (CH)$_x$
in the ideally dimerized zig-zag Lewis structure 
with alternating double and single bonds between the carbon atoms (black dots).
Linked to each carbon atoms are the hydrogen atoms (white circles)
on alternating sides of the chain.\label{fig:structure}}
\end{figure}

The H\"uckel--Hubbard-Ohno Hamiltonian includes the Coulomb interaction
of the $\pi$-electrons in the Hubbard-Ohno approximation,
\begin{eqnarray}
\hat{H}^{\rm HHO}&=& \hat{H}^{\rm H}
+U\sum_{l}\left(\hat{n}_{l,\uparrow}-1/2\right)\left(\hat{n}_{l,\downarrow}-1/2\right)
\nonumber \\
&& + \frac{1}{2\epsilon_d}\sum_{l\neq m}V_{l,m}
\left(\hat{n}_l-1\right)\left(\hat{n}_m-1\right)
\end{eqnarray}
with the H\"uckel part $\hat{H}^{\rm H}$ from eq.~(\ref{eq:Hueckelhamilt})
and
\begin{equation}
V_{l,m}=V(|\vec{r}_l-\vec{r}_m|)
\end{equation}
with $V(x)$ from eq.~(\ref{eq:Ohnoexpression}).
The carbon atoms are at the positions~$\vec{r}_l$ of a zig-zag chain
in the $x$-$y$-plane as is shown in Fig.~\ref{fig:structure}.
We optimize such trans-polyacetylene geometry numerically as will
be discussed briefly below.

\subsubsection{Numerical procedure}

We employ the density-matrix renormalization group (DMRG)
method~\cite{White-1992,White-1993} 
together with the dynamic block state selection (DBSS) 
approach,~\cite{Legeza-2003,Legeza-2004}
to calculate the ground-state energy and excited-state energies
for the H\"uckel--Hubbard-Ohno model for up to $L_{\rm C}=66$ carbon atoms
($L_C=2L+2$) for fixed bond parameters. We use a fixed-point recursion method to 
optimize the geometry.~\cite{BarfordPDA} In this procedure, DMRG is called
iteratively, thus playing the role of
a kernel function. The electronic Hamiltonian for a fixed geometry
is solved by the DMRG, and expectation values of the required operators
are determined from the obtained wavefunction. These quantities are used
to minimize the total energy comprising 
the electronic contribution and the lattice potential energy.
This geometrical optimization terminates when the energy difference between 
two subsequent iteration steps gets below an a-priori defined threshold 
that we set to $10^{-4}\, {\rm eV}$.
A more detailed description will be given in a subsequent paper.~\cite{secondpaper}

In addition, we employ the force field
\begin{equation}
F_{\sigma}(r)=-[K_{\sigma,0}+K_{\sigma,1}(r-r_0)] \; ,
\end{equation}
and start our analysis with the parameter set motivated
in Sect.~\ref{sec:HHOmodel},
$K_{\sigma,0}=-4.7\, {\rm eV}/\hbox{\AA}$
and $K_{\sigma,1}=50\, {\rm eV}/\hbox{\AA}{}^2$.
Moreover, we show data for an improved parameter set,
$K_{\sigma,0}=-4.8\, {\rm eV}/\hbox{\AA}$
and $K_{\sigma,1}=42\, {\rm eV}/\hbox{\AA}{}^2$.
For a comparison with our earlier study on polydi\-acetylene, we fix
$U=6\, {\rm eV}$ and $V=3\, {\rm eV}$. Furthermore, the 
dielectric screening of the surrounding medium is incorporated 
using the dielectric constant $\epsilon_d=2.3$ for the long-range part
of the Coulomb interaction.
Moreover, we use $t_0=2.5\, {\rm eV}$, $\alpha=4.0\, {\rm eV}$.

\subsubsection{Bond lengths}

\begin{figure}[tb]
\hspace*{-24pt}
\includegraphics[width=7.5cm]{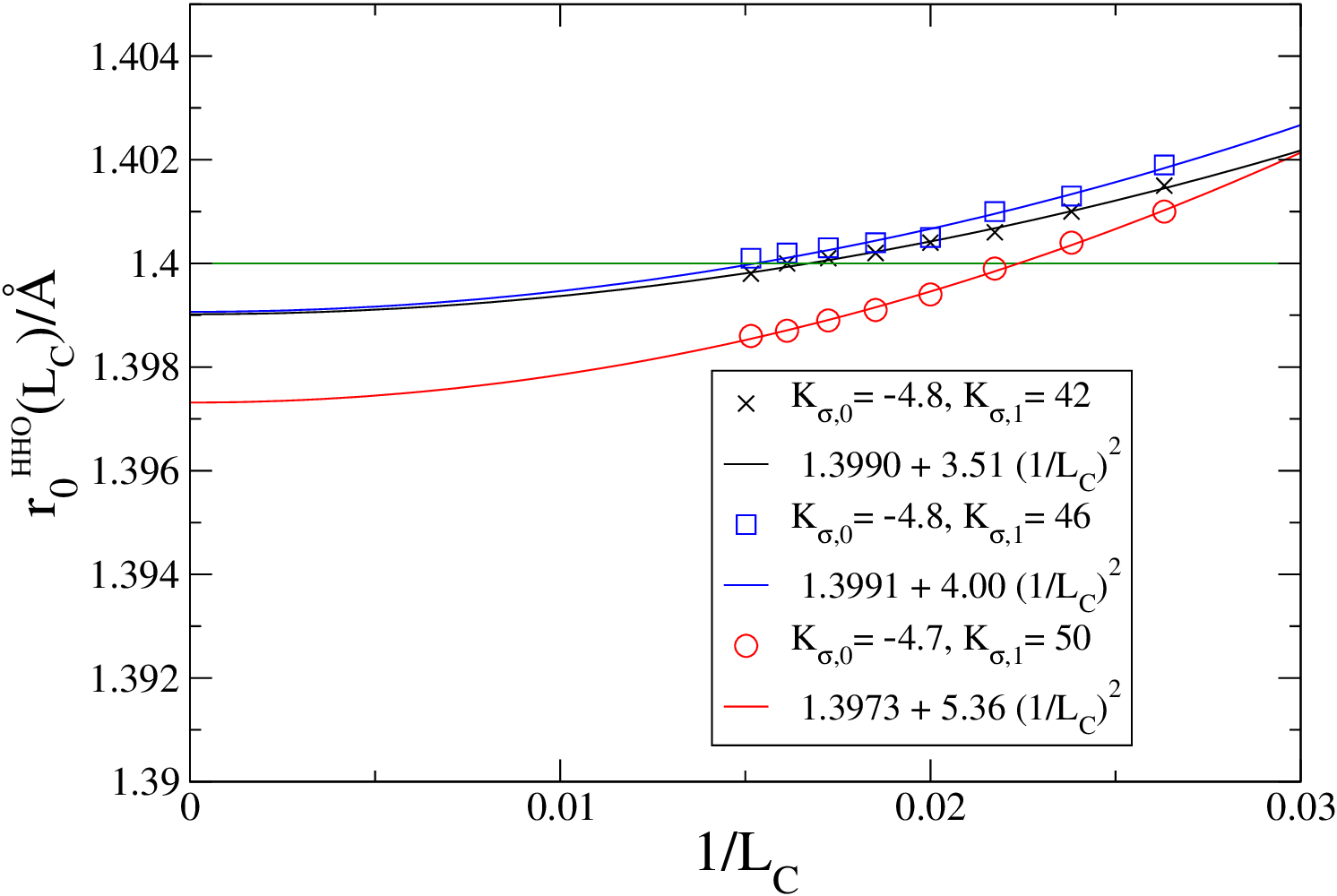}
\caption{(Color online) 
Average bond length $r_0^{\rm HHO}(L_C)$ in Angstr\o m 
as a function of the inverse system size~$1/L_C$
in the H\"uckel--Hubbard-Ohno model with $\alpha=4.0\, {\rm eV}/\hbox{\AA}$,
$t_0=2.5\, {\rm eV}$, and a linear $\sigma$-bond force field $F_{\sigma}(r)$
with $K_{\sigma,0}=-4.7\, {\rm eV}/\hbox{\AA}$,
$K_{\sigma,1}=50\, {\rm eV}/\hbox{\AA}{}^2$ (red circles),
$K_{\sigma,0}=-4.8\, {\rm eV}/\hbox{\AA}$,
$K_{\sigma,1}=46\, {\rm eV}/\hbox{\AA}{}^2$ (blue squares),
and
$K_{\sigma,0}=-4.8\, {\rm eV}/\hbox{\AA}$,
$K_{\sigma,1}=42\, {\rm eV}/\hbox{\AA}{}^2$ (black crosses).
The experimental value $r_0=1.40\, \hbox{\AA}$ is shown by a horizontal line,
the quadratic extrapolation lines are included as a guide to the eye.
\label{fig:DMRG-bonds}}
%
\begin{center}
\mbox{}\\[9pt]
\hspace*{-24pt}
\includegraphics[width=7.5cm]{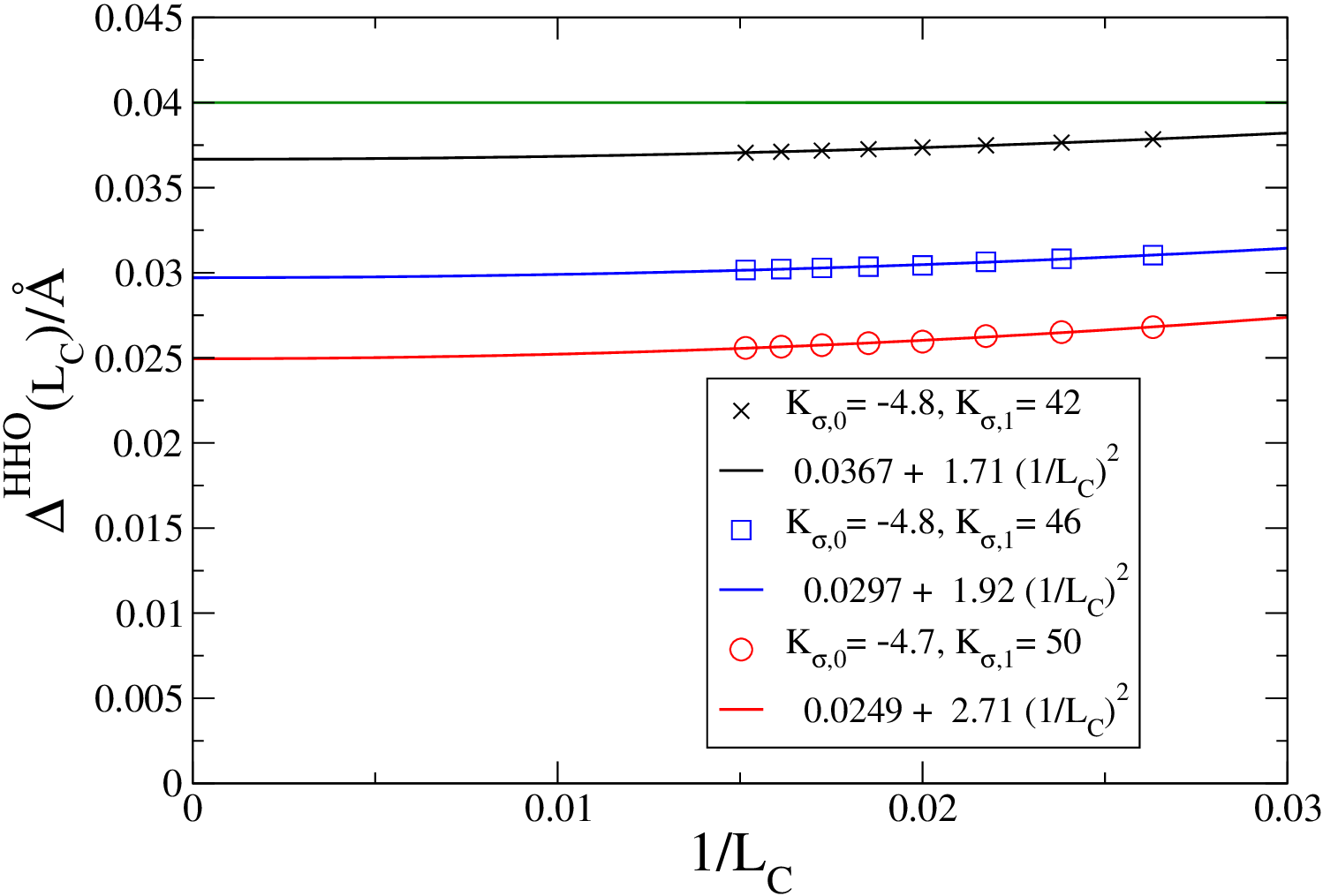}
\end{center}
\caption{(Color online) 
Dimerization~$\Delta_0^{\rm HHO}(L_C)$ in Angstr\o m 
as a function of the inverse system size~$1/L_C$
in the H\"uckel--Hubbard-Ohno model with $\alpha=4.0\, {\rm eV}/\hbox{\AA}$,
$t_0=2.5\, {\rm eV}$, and a linear $\sigma$-bond force field $F_{\sigma}(r)$
with $K_{\sigma,0}=-4.7\, {\rm eV}/\hbox{\AA}$,
$K_{\sigma,1}=50\, {\rm eV}/\hbox{\AA}{}^2$ (red circles),
$K_{\sigma,0}=-4.8\, {\rm eV}/\hbox{\AA}$,
$K_{\sigma,1}=46\, {\rm eV}/\hbox{\AA}{}^2$ (blue squares),
and
$K_{\sigma,0}=-4.8\, {\rm eV}/\hbox{\AA}$,
$K_{\sigma,1}=42\, {\rm eV}/\hbox{\AA}{}^2$ (black crosses).
The experimental value~$\Delta_0=0.04\, \hbox{\AA}$ 
is shown as a horizontal line,
the quadratic extrapolation lines are included as a guide to the eye.
\label{fig:DMRG-dimer}}
\end{figure}

The results for the average bond length~$r_0^{\rm HHO}$ 
are shown in Fig.~\ref{fig:DMRG-bonds}.
As for the non-interacting case, the H\"uckel--Hubbard-Ohno model
almost perfectly describes the bond-length reduction
from $r_{\sigma}$ to $r_0$. 
We find
$r_0^{\rm HHO,1}=1.397\, \hbox{\AA}$ 
for our first parameter set ($K_{\sigma,0}=-4.7\, {\rm eV}/\hbox{\AA}$,
$K_{\sigma,1}=50\, {\rm eV}/\hbox{\AA}{}^2$) 
as motivated in Sect.~\ref{sec:HHOmodel},
and 
$r_0^{\rm HHO,2}=1.399\, \hbox{\AA}$ 
for the improved, second parameter set ($K_{\sigma,0}=-4.8\, {\rm eV}/\hbox{\AA}$,
$K_{\sigma,1}=42\, {\rm eV}/\hbox{\AA}{}^2$),
in very good agreement with experiment, $r_0=1.40\, \hbox{\AA}$. 
Now that $r_0^{\rm H,1}=r_0^{\rm H,2}=1.391\, \hbox{\AA}$ 
for $\alpha=4.0\, {\rm eV}/\hbox{\AA}$, 
we see that
the electron-electron interaction and the parameter $K_{\sigma,1}$ do not 
significantly influence the average bond length.
Moreover, the analysis of the
bond potential $V_{\sigma}(r)$ provides a good estimate of $K_{\sigma,0}$.

In contrast, the dimerization sensitively depends on the parameter for 
the $\sigma$-bond spring constant $K_{\sigma,1}$, see Fig.~\ref{fig:DMRG-dimer}.
Moreover, the estimate for $K_{\sigma,1}=50\, {\rm eV}/\hbox{\AA}{}^2$
as motivated in Sect.~\ref{sec:HHOmodel} leads to a too small dimerization even
in the presence of the electron-electron interaction.
Indeed, we find
$\Delta_0^{\rm HHO,1}=0.025\, \hbox{\AA}$ 
for the parameter set ($K_{\sigma,0}=-4.7\, {\rm eV}/\hbox{\AA}$,
$K_{\sigma,1}=50\, {\rm eV}/\hbox{\AA}{}^2$),
as compared to $\Delta_0^{\rm H,1}=0.006\, \hbox{\AA}$
from the bare H\"uckel model and $\Delta_0=0.04\, \hbox{\AA}$ from experiment.
Apparently, the electron-electron interaction substantially increases
the dimerization. 
For the first parameter set, the dimerization in presence of the
Hubbard-Ohno interaction is by a factor five larger than the Peierls 
contribution alone.

\begin{figure}[b]
\begin{center}
\mbox{}\\[18pt]
\hspace*{-24pt}
\includegraphics[width=8cm]{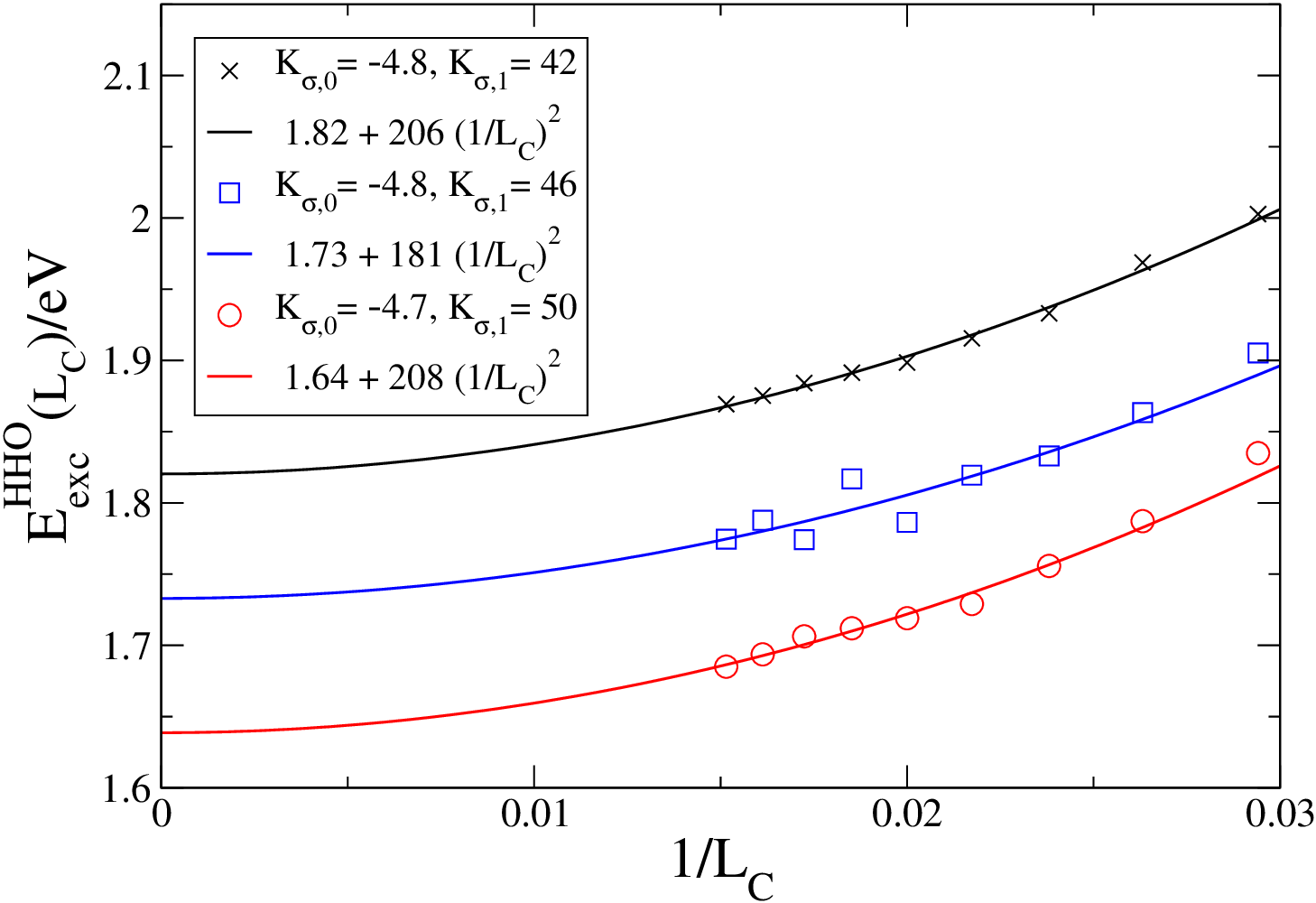}
\end{center}
\caption{(Color online) 
Exciton energy $E_{\rm exc}(L_C)$ in eV
as a function of the inverse system size~$1/L_C$
in the H\"uckel--Hubbard-Ohno model with $\alpha=4.0\, {\rm eV}/\hbox{\AA}$,
$t_0=2.5\, {\rm eV}$, and a linear $\sigma$-bond force field $F_{\sigma}(r)$
with $K_{\sigma,0}=-4.7\, {\rm eV}/\hbox{\AA}$,
$K_{\sigma,1}=50\, {\rm eV}/\hbox{\AA}{}^2$ (red circles),
$K_{\sigma,0}=-4.8\, {\rm eV}/\hbox{\AA}$,
$K_{\sigma,1}=46\, {\rm eV}/\hbox{\AA}{}^2$ (blue squares),
and
$K_{\sigma,0}=-4.8\, {\rm eV}/\hbox{\AA}$,
$K_{\sigma,1}=42\, {\rm eV}/\hbox{\AA}{}^2$ (black crosses).
The quadratic extrapolation lines are included as a guide to the eye.
\label{fig:DMRG-exc}}
\end{figure}

To obtain a better agreement with the experimentally observed dimerization,
we address the improved parameter set ($K_{\sigma,0}=-4.8\, {\rm eV}/\hbox{\AA}$,
$K_{\sigma,1}=42\, {\rm eV}/\hbox{\AA}{}^2$).
We find $\Delta_0^{\rm HHO,2}=0.037\, \hbox{\AA}$,
as compared to $\Delta_0^{\rm H,2}=0.016\, \hbox{\AA}$ and
$\Delta_0=0.04\, \hbox{\AA}$. The value for the dimerization now
agrees with the experimental value within experimental error bars.
The electron-electron interaction still is decisive for the dimerization
in trans-polyacetylene because the electronic contribution to the dimerization 
is a factor 1.4 larger than the Peierls contribution.

\subsubsection{Band gaps}

\begin{figure}[tb]
\begin{center}
\mbox{}\\[9pt]
\hspace*{-24pt}
\includegraphics[width=8cm]{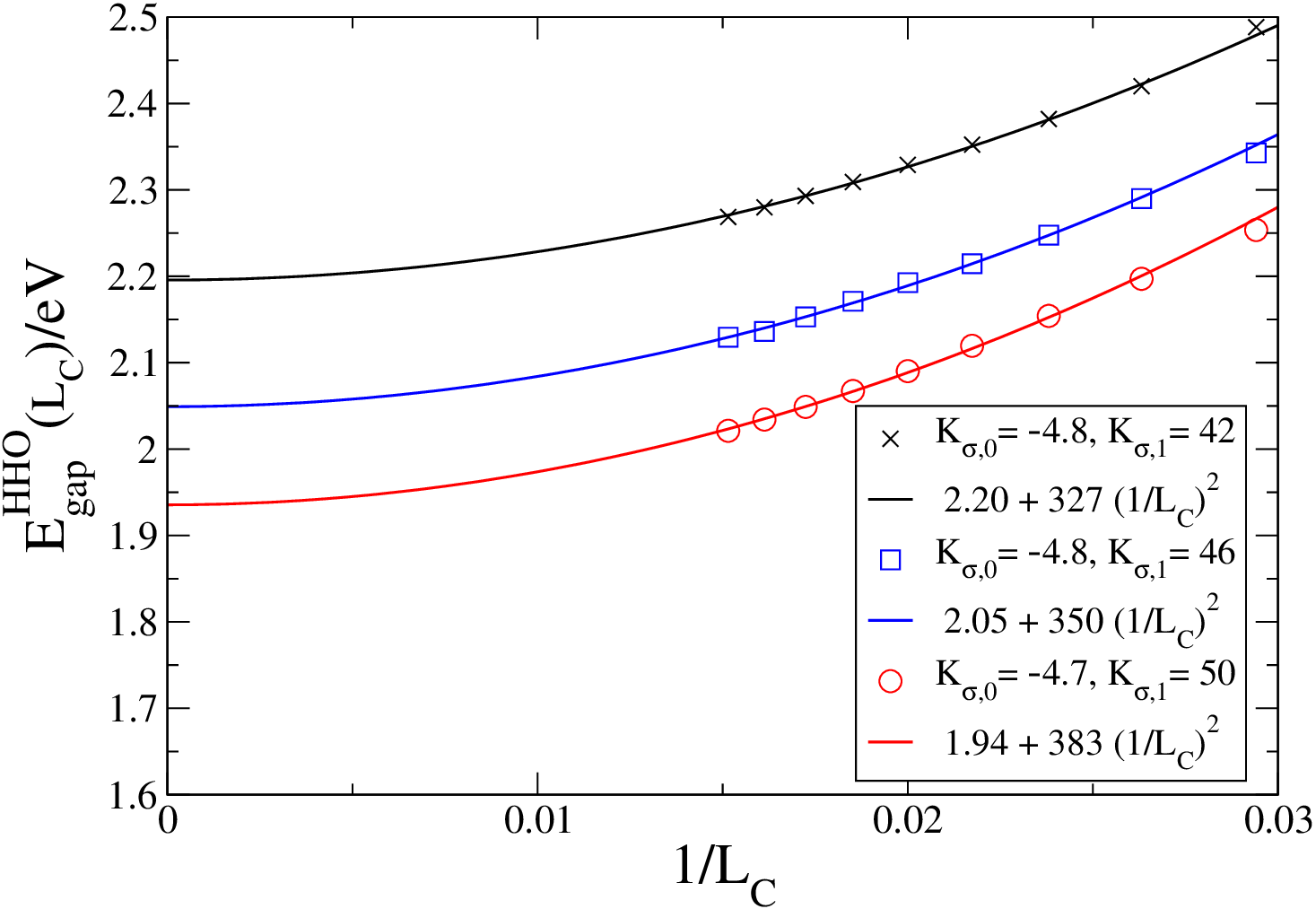}
\end{center}
\caption{(Color online) 
Single-particle gap $E_{\rm gap}(L_C)$ in eV
as a function of the inverse system size~$1/L_C$
in the H\"uckel--Hubbard-Ohno model with $\alpha=4.0\, {\rm eV}/\hbox{\AA}$,
$t_0=2.5\, {\rm eV}$, and a linear $\sigma$-bond force field $F_{\sigma}(r)$
with $K_{\sigma,0}=-4.7\, {\rm eV}/\hbox{\AA}$,
$K_{\sigma,1}=50\, {\rm eV}/\hbox{\AA}{}^2$ (red circles),
$K_{\sigma,0}=-4.8\, {\rm eV}/\hbox{\AA}$,
$K_{\sigma,1}=46\, {\rm eV}/\hbox{\AA}{}^2$ (blue squares),
and
$K_{\sigma,0}=-4.8\, {\rm eV}/\hbox{\AA}$,
$K_{\sigma,1}=42\, {\rm eV}/\hbox{\AA}{}^2$ (black crosses).
The quadratic extrapolation 
lines are included as a guide to the eye.
\label{fig:DMRG-gaps}}
\end{figure}

The H\"uckel--Hubbard--Ohno results for 
the exciton energy and the single-particle gap 
are shown in Figs.~\ref{fig:DMRG-exc} and~\ref{fig:DMRG-gaps},
respectively.

The exciton energies extrapolate to $E_{\rm exc}^{\rm HHO,1}=1.64\, {\rm eV}$
and $E_{\rm exc}^{\rm HHO,2}=1.82\, {\rm eV}$, within the experimental bounds
$1.5\, {\rm eV}< E_{\rm exc}< 1.9\, {\rm eV}$.
The value from the improved parameter set is in good agreement with
the prediction from third-harmonic generation.~\cite{Fann}
The position of the exciton energy sensitively depends on
the force-field parameter~$K_{\sigma,1}$. 
The corresponding results for the single-particle gap are
$E_{\rm gap}^{\rm HHO,1}=1.94\, {\rm eV}$ and
$E_{\rm gap}^{\rm HHO,2}=2.20\, {\rm eV}$.
The binding energy
of the exciton $\Delta_s=
E_{\rm gap}-E_{\rm exc}$ is predicted to be 
$\Delta_s^{\rm HHO,1}= 0.30\, {\rm eV}$, and
$\Delta_s^{\rm HHO,2}= 0.38\, {\rm eV}$,
in agreement with experiment.~\cite{Liess}
Apparently, the binding energy $\Delta_s$ increases slowly with decreasing
force-field parameter~$K_{\sigma,1}$.

\section{Conclusions}
\label{sec:conclusions}

In this work we investigated the $\sigma$-bonds 
and $\pi$-bonds in ethane (H$_3$C$-$CH$_3$),
ethene (a.k.a.\ ethylene, H$_2$C$=$CH$_2$), 
and ethyne (a.k.a.\ acetylene,
HC$\equiv$CH) as a function of the carbon-carbon distance~$r$.
We demonstrated that the $\pi$-bonds 
in ethene and ethyne can be described
using the H\"uckel model with the potential $V_{\sigma}(r)$
from the $\sigma$-bond.

The bond lengths and spring constants
in equilibrium agree with the data from quantum chemistry
within a margin of a few percent.
The comparison provides a robust estimate for 
the value of the Peierls coupling, $\alpha=4\, {\rm eV}/\hbox{\AA}$,
for a given electron transfer $t_0=2.5\, {\rm eV}$ 
at carbon-carbon distance $r_0=1.4\, \hbox{\AA}$.
Unfortunately, the parameters of the
Hubbard-Ohno interaction cannot be determined from 
dimers or short polyenes because the Ohno interaction
is essentially constant for small distances.
Therefore, we choose $U=6\, {\rm eV}$ and $V=3\, {\rm eV}$
as derived from the the analysis of excited states in 
polydiacetylene.~\cite{BarfordPDA}

We tested the H\"uckel--Hubbard-Ohno model for trans-polyacetylene
with $V_{\sigma}(r)$ as backbone potential.
Close to the average bond length~$r_0$, we may
linearize the $\sigma$-bond force $F_{\sigma}(r)=-V_{\sigma}'(r)
=-[K_{\sigma,0}+K_{\sigma,1}(r-r_0)]$ when we determine the structure.
The $\sigma$-bond repulsion balances the lattice contraction induced
by the itinerant $\pi$-electrons.
We find that the size of the average bond contraction is mostly
determined by the constant term $K_{\sigma,0}$ and the Peierls coupling~$\alpha$
whereas the Coulomb interaction and the linear term $K_{\sigma,1}$
are fairly unimportant for the average bond length.
Moreover, the optimal value for $K_{\sigma,0}$
is very well predicted by $V_{\sigma}(r)$, 
$K_{\sigma,0}^{(1)}=-5.2\, {\rm eV}/\hbox{\AA}$ while
our analysis for trans-polyacetylene suggests
$K_{\sigma,0}^{(2)}=-4.8\, {\rm eV}/\hbox{\AA}$.

The dimerization~$\Delta_0$ in trans-polyacetylene
is triggered to a large part by the Coulomb interaction.
Naturally, the dimerization strength sensitively depends 
on the size of the `spring constant' $K_{\sigma,1}$. 
We find that the value obtained from
the analysis of the $\sigma$-bond potential,
$K_{\sigma,1}^{(1)}=54\, {\rm eV}/\hbox{\AA}{}^2$, is too large.
For an agreement with experimental data in trans-polyacetylene,
we propose to use the smaller value
$K_{\sigma,1}^{(2)}=42\, {\rm eV}/\hbox{\AA}{}^2$,
in agreement with empirical values for benzene,
$K_{\sigma}^{\rm ben}=41.3\, {\rm eV}/\hbox{\AA}{}^2$,
that has the same average bond length.~\cite{benzene}   
Our value is considerably larger than 
$K_{\sigma}^{{\rm PA},1}=31\, {\rm eV}/\hbox{\AA}{}^2$
proposed for trans-polyacetylene in Ref.~[\onlinecite{smallKinPA}]
and closer to the value suggested 
in Ref.~[\onlinecite{largeKinPA}],
$K_{\sigma}^{{\rm PA},2}=46\, {\rm eV}/\hbox{\AA}{}^2$.

Given these parameter sets, we calculated the energy for
elementary excitations in trans-polyacetylene.
We find an exciton with substantial binding energy,
$\Delta_s\approx 0.4\, {\rm eV}$, in agreement with experiments
for $\pi$-conjugated materials.~\cite{Liess}
The parameter $K_{\sigma,1}$ mildly affects
the exciton binding energy 
but determines the energetic position of the exciton.
For our second, optimal parameter set [$t_0=2.5\, {\rm eV}$,
$\alpha=4.0\, {\rm eV}/\hbox{\AA}$,
$U=6\, {\rm eV}$, $V=3\, {\rm eV}$,
$K_{\sigma,0}^{(2)}=-4.8\, {\rm eV}/\hbox{\AA}$,
$K_{\sigma,1}^{(2)}=-42\, {\rm eV}/\hbox{\AA}{}^2$],
we find a good agreement with experiment,
$E_{\rm exc}=1.8 \, {\rm eV}$.\cite{Fann,Leising}
The single-particle gap is found at $E_{\rm gap}=2.2\, {\rm eV}$.
Our values for $E_{\rm exc}$ and $E_{\rm gap}$
also agree with the predictions 
from recent calculations of Mott-Wannier excitons
for correlated electrons in one dimension.~\cite{Rohlfing}

In sum, the analysis of short molecules gives a reasonable
first estimate for the parameters necessary for a H\"uckel--Hubbard-Ohno
description of conjugated polymers. To reproduce
the experimental ground-state conformation, 
some parameters, typically the `spring constant' at the optimal
average bond length, must be adjusted by some 20\%. 
After the adjustment of these parameters, the 
H\"uckel--Hubbard-Ohno model provides a fairly
good description for $\pi$-electrons in trans-polyacetylene,
as can be seen from the good agreement
of theoretical and experimental data for
the exciton energy and the exciton binding energy.

The measured optical phonon spectra provide another testing case
for the H\"uckel--Hubbard-Ohno model.
Indeed, this theoretical approach reproduces
the optical phonons in trans-polyacetylene with good accuracy,
as we shall show in a subsequent paper.~\cite{secondpaper}

\begin{acknowledgments}
We thank  G.\ Frenking, M.\ Hermann, and P.\ Knowles 
for useful discussions and for providing us their
quantum-chemistry data for comparison.
This research was supported in part by the Hungarian Research Fund (OTKA) under
Grant Nos.~K~100908 and NN~110360.
L.V.\ was supported by the Grant Agency of the Czech Republic 
(Grant No~15-10279Y and Grant No.~16-12052S).
\end{acknowledgments}



\begin{thebibliography}{99}

\bibitem{Jensen} F.~Jensen, {\sl Introduction to Computational Chemistry}
(Wiley, Chichester, 1999).

\bibitem{Frenking} {\sl The Chemical Bond}, ed.\ by G.~Frenking and S.~Shaik,
vol.~1 and~2 (Wiley-VCH, Weinheim, 2014).

\bibitem{Baeriswyl} D.~Baeriswyl in {\sl Conjugated Conducting Polymers},
ed.\ by H.\ G.\  Kiess (Springer Series in Solid-State Sciences~{\bf 102},
Springer, Berlin, Heidelberg, 1992), p.~7.

\bibitem{Barford} W.~Barford, {\sl Electronic and Optical Properties
of Conjugated Polymers} (Clarendon Press, Oxford, 2005, 2009).

\bibitem{BarfordPDA} 
G.\ Barcza, W.\ Barford, F.\ Gebhard, and \"O.\ Legeza, 
Phys.\ Rev.\ B~{\bf 87}, 245116 (2013).

\bibitem{SSHdebate} S.\ Kivelson, W.-P.\ Su, J.\ R.\ Schrieffer, and A.\ J.\ Heeger,
Phys.\ Rev.\ Lett.~{\bf 58}, 1899 (1987);
D.\ Baeriswyl, P.\ Horsch, and K.\ Maki,
Phys.\ Rev.\ Lett.~{\bf 60}, 70 (1987);
J.\ T.\ Gammel and D.\ K.\ Campbell,
Phys.\ Rev.\ Lett.~{\bf 60}, 71 (1987);
S.\ Kivelson, W.-P.\ Su, J.\ R.\ Schrieffer, and A.\ J.\ Heeger,
Phys.\ Rev.\ Lett.~{\bf 60}, 72 (1987).

\bibitem{Gammel} D.\ K.\ Campbell, J.\ T.\ Gammel, and E.\ Y.\ Loh,
Phys.\ Rev.\ B~{\bf 42}, 475 (1990).

\bibitem{smallKinPA} A.\ Girlando, A.\ Painelli, G.\ W.\ Hayden, and Z.\ G.\ Soos,
Chem.\ Phys.~{\bf 184}, 139 (1994).

\bibitem{Herzberg} G.~Herzberg, 
{\sl Electronic spectra and electronic structure of polyatomic molecules}
(Van Nostrand, New York, 1966).

\bibitem{Helvoort} K.\ van Helvoort, W.~Knippers,
R.\ Fantoni, and S.\ Stolte, Chem.\ Phys.~{\bf 111}, 445 (1987).

\bibitem{largeKinPA} E.\ Ehrenfreund, Z.\ Vardeny, O.\ Brafman, and B.\ Horovitz,
Phys.\ Rev.\ B~{\bf 36}, 1535 (1987).

\bibitem{benzene} 
L.\ Goodman, A.\ C.\ Ozkabak, and S.\ N.\ Thakurt,
J.\ Phys.\ Chem.~{\bf 95}, 9044 (1991);
S.\ Rashev and D.\ C.\ Moule,
J.\ Phys.\ Chem.\ A~{\bf 108}, 1259 (2004).

\bibitem{Mele} E.\ J.\ Mele, Mol.\ Chryst.\ Liq.\ Chryst.~{\bf 77}, 25 (1981).

\bibitem{NMR} C.\ S.\ Yannoni and T.\ C.\ Clarke, Phys.\ Rev.\ Lett.~{\bf 51}, 
1191 (1983).

\bibitem{MolPro} H.-J.\ Werner, P.\ J.\ Knowles, G,\ Knizia, F.\ R.\
Manby, and M.\ Sch\"utz,
WIREs Comput.\  Mol.\ Sci.~{\bf 2}, 242 (2012).

\bibitem{basis} T.\ H.\ Dunning, Jr., J.\ Chem.\ Phys.~{\bf 90}, 1007
(1989).

\bibitem{Knowles} J.\ B.\ Robinson and P.\ J.\ Knowles, J.\ Chem. Theory and 
Comp.~{\bf 8}, 2653 (2012).

\bibitem{Frenkingdata} M.\ Hermann and G.\ Frenking, 
submitted to Chemistry -- A European Journal (2015).

\bibitem{LegezaMottet}  M.\ Mottet, P.\ Tecmer, K.\ Boguslawski, 
\"O.\ Legeza, and M.\ Reiher, Phys.\ Chem.\ Chem.\ Phys.~{\bf 16}, 8872 (2014).

\bibitem{Lievin} J.\ Li\`{e}vin, J.\ Demaison, M.\ Herman, A.\ Fayt, 
and C.\ Puzzarini,
J.\ Chem.\ Phys.~{\bf 134}, 064119 (2011).

\bibitem{Krasser} W.\ Krasser, A.\ Fadini, E.\ Rozemuller, and A.\ J.\ 
Renou\-prez, J.\ Mol.\ Struc.~{\bf 66}, 135 (1980).

\bibitem{FastWelsh} H.\ Fast and H.\ L.\ Welsh, 
J.\ Mol.\ Spec.~{\bf 41}, 203 (1972).

\bibitem{Lepetit} M.-B.\ Lepetit and G.\ M.\ Pastor,
Phys.\ Rev.\ B~{\bf 56}, 4447 (1997).

\bibitem{Chinesenfit} D.\ Zhang, Q.\ Qu, C.\ Liu, and Y.\ Jiang,
J.\ Chem.\ Phys.~{\bf 134}, 024114 (2011).

\bibitem{SSH} A.\ J.\ Heeger, S.\ Kivelson,
J.\ R.\ Schrieffer, and W.-P.\ Su,
Rev.\ Mod.\ Phys.~{\bf 60}, 781 (1988).

\bibitem{PhilMag} F.\ Gebhard, K.\ Bott, M.\ Scheidler, P.\ Thomas, 
and S.\ W.\ Koch,
Phil.\ Mag.\ B~{\bf 75}, 1 (1997).

\bibitem{Mathematica} Mathematica, Version 9.0 (Wolfram Research, Inc., 
Champaign, 2012).

\bibitem{Orenstein} J.\ Orenstein, G.\ L.\ Baker, and Z.\ Vardeny,
J. de Phys.\ Colloques~{\bf 44}, C3-407 (1983).

\bibitem{Phillips} S.\ D.\ Phillips, R.\ Worland, H.\ Yu, T.\ Hagler, R.\ Freedman,
Y.\ Cao, V.\ Yoon, J.\ Chiang, W.\ C.\ Walker, and A.\ J.\ Heeger,
Phys.\ Rev.\ B~{\bf 40}, 9751 (1989).

\bibitem{Fann} 
W.-S.\ Fann, S.\ Benson, J.\ M.\ J.\  Madey,
S.\ Etemad, G.\ L.\ Baker, and F.\ Kajzar,
Phys.\ Rev.\ Lett.~{\bf 62}, 1492 (1989).

\bibitem{Leising} G.\ Leising, Phys.\ Rev.\ B~{\bf 38}, 10313 (1988).

\bibitem{Kubo} T.\ Kubo, T.\ Watanabe, T.\ Nishioka, H.\ Takezoe, and A.\ Fukuda,
Jpn.\ J.\ Appl.\ Phys.~{\bf 31}, 3372 (1992).

\bibitem{Liess} M.\ Liess, S.\ Jeglinski, Z.\ V.\ Vardeny, M.\ Ozaki, K.\ Yoshino,
Y. \ Ding and T.\ Barton, Phys.\ Rev.\ B~{\bf 56}, 15712 (1997).

\bibitem{Rohlfing} M.\ Rohlfing and S.\ G.\ Louie,
Phys.\ Rev.\ Lett.~{\bf 82}, 1959 (1999).

\bibitem{White-1992}
S.~R.~White, Phys.~Rev.~Lett. {\bf 69},  2863  (1992).

\bibitem{White-1993}
S.~R.~White, Phys.~Rev.~B {\bf 48},  10345  (1993).
 
\bibitem{Legeza-2003}
\"O. Legeza, J. R\"oder, and B. A. Hess, Phys. Rev. B {\bf 67}, 125114 (2003).

\bibitem{Legeza-2004}
\"O. Legeza and J. S\'olyom, Phys. Rev. B {\bf 70}, 205118 (2004).

\bibitem{secondpaper} M.\ Tim\'ar, G.\ Barcza, \"O.\ Legeza, and
F.\ Gebhard, in preparation.


\end{thebibliography}
\end{document}